\documentclass[pre,reprint,nofootinbib,superscriptaddress]{revtex4-2}
\pdfoutput=1

\usepackage[colorlinks,linkcolor=blue,urlcolor=blue,citecolor=blue]{hyperref}
\usepackage{physics}

\usepackage[english]{babel}

\usepackage{xcolor}
\usepackage{xspace}
\usepackage{ifthen}

\begin{document}

\title{Quanta Iff Discreteness}

\author{Marcello {Poletti}}
\email{epomops@gmail.com}
\affiliation{San Giovanni Bianco, Italy}

 \begin{abstract}
A brief philosophical inquiry into the foundations of quantum mechanics is presented here. In particular, the direct relationship between granularity, discontinuity, and the presence of quantum effects will be argued. Furthermore, an "interpretation of relational interpretation" will be supported, which, in combination with the problem of logical undecidability, produces a promising approach that places the apparent illogicality of QM within the realm of logic and effectively addresses its usual paradoxes.

\end{abstract}

\maketitle

\section{Presentation}
Quantum mechanics arose as a theory of discrete phenomena, rooted in Planck's blackbody radiation problem \cite{Planck}, Einstein's photoelectric effect \cite{Einstein}, and Bohr's atomic model \cite{Bohr}.

Building upon these foundations, Heisenberg, Born, and Jordan developed matrix mechanics \cite{Heis1, BoJo, BoJoHe}, a discrete mechanics in which the possible discrete values of observable quantities are placed in the discrete cells of matrices.

The resulting theory proves to be extraordinarily effective and rich in unexpected effects, such as the principle of superposition, the uncertainty principle, non-commutativity, the tunneling effect, the violation of Bell's inequalities, the (presumed) non-locality, and so on. These effects, in many surprising and paradoxical ways, surpass the semiclassical origins of QM, and the discrete nature of reality becomes somewhat secondary.

In fact, the presence of discrete phenomena does not seem to have any inherent relationship with the peculiarities of quantum mechanics. 

If we consider any strongly discrete model, such as a cellular automaton like Game of Life \cite{Gardner}, it does not exhibit any quantum behavior; the automaton evolves according to classical rules without raising any of the epistemological and ontological problems triggered by QM. But are we really sure of this?

An automaton cellular is a game observed by a deus ex machina who studies its evolution; in this sense, it is more of a metaphysical\footnote{By the term "metaphysics" we mean here the study of reality in itself, not subject to the constraint of the scientific method.} model than a physical one. The question that should be asked in a case like this is more subtle, and the answer is decidedly less obvious: would a "sentient being"\footnote{It is not intended to give any specific value to consciousness here; the use of the term "sentient being" is purely rhetorical.} living within that game describe the world around them with a classical theory? Or would they observe quantum phenomena? The basic philosophical point, if you will, is exquisitely relational \cite{Rovelli}. Physics is a \textbf{description} of the world from a viewpoint internal to the world. Being a discipline constrained by the scientific method, it is verified by the outcomes of concrete measurements on things in the world and therefore has a dual constraint in that it deals only with what one part of the world can measure about another part of the world and the possible outcomes of these measurements.

Consider two classical, Newtonian bodies, A and B, subject to mutual gravitational attraction. The dynamics of this system are given by writing a certain Lagrangian and setting the initial conditions. Once this is done, the history of the system is fixed and can be calculated more or less simply depending on the more or less complex form of the Lagrangian.

Once again, however, this simple approach has a metaphysical flavor as it is entirely unrealistic. It is not possible to observe two solitary bodies in a classical space-time; if the observer is there to make measurements, their own gravitational field will also intervene, and the entire system will be altered. This is nothing but the well-known effect of the unavoidable disturbance that the observer causes to the observed, as emphasized by Heisenberg in his philosophical writings \cite{Heis2, Laudisa} to justify his uncertainty principle.

However, in the classical context, this effect is \textbf{negligible}. Let us suppose a point of view that coincides with body A (or equivalently B), then it is legitimate to ask how A describes the world, and nothing changes. The observer A will set the Lagrangian, the initial conditions, and calculate the dynamics of the system, including itself, without the need for an external observer/disturber. In this way, any annoying observations of philosophical principle are circumvented.

However, the problems are not entirely eliminated, and it is necessary to ask whether A can perform the required operations, namely write the Lagrangian and set the initial conditions. Let us focus on this second point. Body A must be able to determine its own distance from B and relative velocity. To do so, it cannot use a tool such as a ruler or a laser measurer or any other, as this would introduce disturbance elements into the system again. Nevertheless, A can measure its own acceleration (in the direction of B) by acquiring essential information from a simple local measurement.

This is a magical effect of classical physics, where \textit{local observations allow for the deduction of complete information about remote systems}. We can look up at the sky and discover the existence of stars without any interaction with them, simply by analyzing some photons that are here, around us. A mirror vibrates in a tunnel, and we know that two black holes have rotated around each other in a place and time remote from us \cite{Ligo}. This is what physics is about, not Kantian noumena \cite{Kant}, not reality itself, but rather phenomena, our sensory interactions, our local space-time. Physics deals with the ability to deduce from local phenomena, the states of remote reality.

Classical physics is magical in that it implicitly postulates (and is consequently formalized) that this transfer of knowledge, of information is absolute.

The masses of the black holes in the LIGO experiment are expressed in solar masses, but classical mechanics and relativity are built in such a way that, in principle, such masses could be determined with arbitrary precision. The noumenon transfers \textbf{all} information continuously into the phenomenon, so much so that noumenon and phenomenon can overlap.

This observation triggers a tension with Occam's razor and with Leibniz's problem of indiscernibles \cite{Leibniz}: if noumenon and phenomenon coincide, what is the point of noumenon? And this leads spontaneously to the most radical solipsism, where my Cartesian ego \cite{Cartesio} is sufficient to describe the entire universe since all the information of the world resides within it.

But let us return to the bodies A and B, what, in classical formalism, ensures that A has all the information about B? It is a deep and absolute continuity between noumenon and phenomenon which is formally translated into a rigorous separation of paths in the phase space of classical physics.

Even more formally, this is guaranteed by the exclusive use, implicitly in classical physics, of analytic functions (see in this regard the beautiful chapter on Eulerian functions in Penrose \cite{Penrose}).

There is, of course, no inherent constraint of this kind. Lagrangians are a priori arbitrary functions. However, if a Lagrangian is not absolutely continuous, for example, if it has a nth derivative that is discontinuous, then the problem of deus ex machina would arise again. One could still calculate the dynamics of the system, but body A would not have absolute knowledge of the entire system, and therefore the external observer would become ineliminable, rendering physics metaphysical and incompatible with the scientific method.

This does not mean that A and B cannot produce good physics, but it is necessary to recognize that this physics may also be very different from the physics of the deus ex machina. It is also not possible to exclude the possibility that A and B may produce excellent metaphysics, intuiting the point of view of the deus ex machina, but their ability to perform measurements will be constrained by their being part of the world.

Hawking, a great provocateur as known, declared philosophy dead \cite{Hawking}, exaggerating a suspicious attitude towards this discipline widespread among physicists, especially Anglo-Saxons.

This suspicion, I believe, is based on a simple and strong argument: \textit{the world is what it is and is not constrained by the whims of philosophers.}

The argument is strong, sensible and agreeable, but it overlooks a critical point. Physics is a \textbf{description} of the world, and as such, it is subject not only to what the world is but also to the rules that govern the description itself (logic foremost). Philosophy certainly needs to avoid arbitrary metaphysical descriptions (the world supported by a turtle supported by another turtle), and the scientific method acts as a filter to prevent this from happening, but physics cannot expect to solve its own problems without deep philosophical reflection, for instance, on the costs that this scientific method entails. The tension between the two disciplines continues to be a key element of the development of both.

Returning to the topic of this writing, what happens in A if the breakdown of continuity results in the inability to have all the information about B? A "quantum fact" happens: A is unable to assign some properties, some elements of reality, to B. The physics expressed by A will be probabilistic, where these probabilities will be triggered by A's inability to assign well-defined states \cite{Landsman} to B, and not by classical mechanisms of random phenomena. The resulting probability theory is different from classical probability theory and coincides with the first postulates of quantum mechanics \cite{Poletti2}.

Note how all of this involves subtle but fundamental changes in perspective and language when approaching the foundational problems of QM.

Firstly, the issue is significantly shifted from the observed to the observer, from the object to the subject (essentially supporting the relational interpretation \cite{Rovelli}). Quantum mechanics does not tell us what reality is like, but rather how the observer of that reality can measure it. The problem is not so much what the actual condition of the cat is \cite{Shro}, but what the actual condition of the experimenter is. It is the experimenter who finds themselves in the position of not being able to attribute certain elements of reality to the cat, not the cat itself that possesses or lacks such elements of reality. Rather, the cat will have similar difficulties attributing certain elements of reality to the experimenter \cite{Poletti1}.

Similarly, the meanings of "ontic" and "epistemological" subtly change.

In the classical analysis of the foundational problems of QM, these two terms are used with a meaning, again, related to the object of observation (the cat). Usually, it is understood that the experimenter is subject to epistemological ignorance if they simply do not know the state of the cat, and is subject to ontic ignorance if the cat is ontologically in an indefinite state.

Now shift the focus from the noumenon to the phenomenon: the experimenter is subject to epistemological ignorance if they simply cannot interpret the information, the phenomena around them, and is subject to ontic ignorance if ontologically such information regarding the cat is not available to them. Again, attention is shifted to the physicist themselves, to the experimenter. It is their problem not being able to attribute a definite state to the cat, not the cat's problem, which can be imagined to be peacefully alive or dead without invoking alternative logics or multiple universes \cite{Everett, Vaidman} or things of that sort. And of course, in the same situation, Wigner's friend can find themselves without triggering any paradox (and no need for a special role of consciousness).

The philosophical framework set forth in the following propositions can be summarized as follows:

\begin{enumerate}
	\item Quantum mechanics is the theory of probabilities applicable to undecided properties \cite{Poletti2}.
	\item A property is undecided for an observer O if, in any way, the transition from noumenon to phenomenon presents a discontinuity, a rupture.
	\item Formally, point 2 is what we call discretization.
\end{enumerate}

Conversely, 3 implies 2, which implies 1.

Two important consequences can be derived from these arguments.

The first consequence is that even though physics is inevitably quantum physics due to some form of granularity in the world, metaphysics may not be quantum, and in fact, it seems that it does not have to be. Einstein's need to understand what the world is and how it works ("God does not play dice") can, in a sense, ignore quantum mechanics. Nevertheless, a concrete physical theory that makes predictions about measurement outcomes, that describes phenomena rather than noumena, will still be subject to the strangeness of QM.

The second consequence is profound and in some aspects mysterious. Realism is consolidated because the possibility of applying Occam's razor in a solipsistic perspective is eliminated. The world is not an extension of my Cartesian ego, because outside of my spatio-temporal environment there is something that I cannot fully describe, something that exists independently of me in a very profound sense. \cite{Poletti1}.It is as if the universe prevents the radical solipsism from being applied, as if the world were the smallest description of itself. Naturally, this reasoning tends to shift the focus back to what the world is, forgetting our relationship with it. From this perspective, without excessively attributing ontology to the world, it is as if the \textbf{description} of the world tends to be essential, respecting the principle of Occam's razor. It is as if once again, from the beginning, it is necessary to reflect on the mysterious and inscrutable opening of the Tractatus: the world is the totality of facts, not of things \cite{Witg}.

The breakdown of continuity, that is, the presence of some discrete phenomenon, is a necessary and sufficient condition for observing quantum phenomena. Ultimately, this is due to the scientific method itself, which requires us to be part of the world and to perform measurements on it.

The scientific method separates physics from metaphysics, but the two branches of knowledge remain strongly interdependent. A good metaphysics provides good indications on how to construct a good physics, and good physics provides good indications on how to construct a good metaphysics.

Feynman has repeatedly stated an aphorism that, given the authority of its author, has become a sort of zeroth law of QM:\textit{"No one understands quantum mechanics"} \cite{Fey}.

Today, one hundred years after the birth of that theory, there are many insights that allow us to reassemble the puzzle from a new and convincing perspective. The relational interpretation has shifted the focus from the ontology of things to the ontology of relations.

The role of logical indecidability \cite{Godel}, already intuited by Wheeler in '74 \cite{Wheeler}, has gradually become consolidated \cite{Szangolies,Brukner,Poletti1, Poletti2}, demonstrating first and foremost that quantum mechanics is not at all illogical.

Perhaps we are approaching the violation of Feynman's rule, and perhaps the circle closes in the most surprisingly banal way possible: the world is quantum because it is not continuous, that is, ultimately, exactly what Planck had already observed in the study of blackbody radiation.

\section{Conclusions}

There is a widely explored parallel between logical undecidability and quantum undecidability.

Consider a logical proposition p that is undecidable in a formal system S.

The problems of quantum mechanics, its paradoxes, its illogicality, \textbf{all} lie in this parallel, in insisting on the question "what is strange about the proposition p?", the answer being "nothing, p has nothing strange about it", it is S that cannot assign a truth value to p, it is the relationship between S and p that has a particularity. In physics, it is necessary to make the same mental leap made by Gödel, separating truth and provability\footnote{In other words, truth is metaphysical and provability is physical.}.

A quantum object itself does not possess anything inherently strange. Its ontology is not more bizarre than that of the observer. It is the relationship between the two that is peculiar: the observed is indecidable relative to the observer. Applying an extension of the language, it becomes undecidable. And how can this occur in the physical world? It happens through a break in the continuity of the transfer of information from the noumenon to the phenomenon.

Is it too simple? Yet, it seems to work well. We have focused for decades on the "true" nature of the electron. But the point is all in our nature, in the nature of our phenomena, of our interactions with the electron. The point is that physics, by construction, does not deal with the nature of the electron but with the measurements that a system S performs on it.

Assuming that there is something sensible in this philosophical analysis, there remains, of course, a problem: what then is the "true" nature of the electron?

I don't know. It's a legitimate metaphysical question that requires a metaphysical answer. Such a metaphysical answer will be nothing different from a usual model (mathematical, geometrical, etc.) and we expect that such a model will exhibit a break in the transfer of information from the noumenon to the phenomenon. It is not necessary for the model to "have quantum characteristics"; such characteristics will emerge spontaneously in the theory of measurement (from physics) that follows from it.

\subsection*{Conflicts of interest.}

The author declare no conflicts of interest.

\nocite{*}

\bibliography{Eng}

\begin{thebibliography}{29}%
\makeatletter
\providecommand \@ifxundefined [1]{%
 \@ifx{#1\undefined}
}%
\providecommand \@ifnum [1]{%
 \ifnum #1\expandafter \@firstoftwo
 \else \expandafter \@secondoftwo
 \fi
}%
\providecommand \@ifx [1]{%
 \ifx #1\expandafter \@firstoftwo
 \else \expandafter \@secondoftwo
 \fi
}%
\providecommand \natexlab [1]{#1}%
\providecommand \enquote  [1]{``#1''}%
\providecommand \bibnamefont  [1]{#1}%
\providecommand \bibfnamefont [1]{#1}%
\providecommand \citenamefont [1]{#1}%
\providecommand \href@noop [0]{\@secondoftwo}%
\providecommand \href [0]{\begingroup \@sanitize@url \@href}%
\providecommand \@href[1]{\@@startlink{#1}\@@href}%
\providecommand \@@href[1]{\endgroup#1\@@endlink}%
\providecommand \@sanitize@url [0]{\catcode `\\12\catcode `\$12\catcode
  `\&12\catcode `\#12\catcode `\^12\catcode `\_12\catcode `\%12\relax}%
\providecommand \@@startlink[1]{}%
\providecommand \@@endlink[0]{}%
\providecommand \url  [0]{\begingroup\@sanitize@url \@url }%
\providecommand \@url [1]{\endgroup\@href {#1}{\urlprefix }}%
\providecommand \urlprefix  [0]{URL }%
\providecommand \Eprint [0]{\href }%
\providecommand \doibase [0]{https://doi.org/}%
\providecommand \selectlanguage [0]{\@gobble}%
\providecommand \bibinfo  [0]{\@secondoftwo}%
\providecommand \bibfield  [0]{\@secondoftwo}%
\providecommand \translation [1]{[#1]}%
\providecommand \BibitemOpen [0]{}%
\providecommand \bibitemStop [0]{}%
\providecommand \bibitemNoStop [0]{.\EOS\space}%
\providecommand \EOS [0]{\spacefactor3000\relax}%
\providecommand \BibitemShut  [1]{\csname bibitem#1\endcsname}%
\let\auto@bib@innerbib\@empty
\bibitem [{\citenamefont {Planck}(1900)}]{Planck}%
  \BibitemOpen
  \bibfield  {author} {\bibinfo {author} {\bibfnamefont {M.}~\bibnamefont
  {Planck}},\ }\bibfield  {title} {\bibinfo {title} {Zur theorie des gesetzes
  der energieverteilung im normalspectrum},\ }\href@noop {} {\bibfield
  {journal} {\bibinfo  {journal} {Verhandlungen der Deutschen Physikalischen
  Gesellschaft}\ }\textbf {\bibinfo {volume} {2}},\ \bibinfo {pages} {202}
  (\bibinfo {year} {1900})}\BibitemShut {NoStop}%
\bibitem [{\citenamefont {Einstein}(1900)}]{Einstein}%
  \BibitemOpen
  \bibfield  {author} {\bibinfo {author} {\bibfnamefont {A.}~\bibnamefont
  {Einstein}},\ }\bibfield  {title} {\bibinfo {title} {On a heuristic viewpoint
  concerning the production and transformation of light.},\ }\href@noop {}
  {\bibfield  {journal} {\bibinfo  {journal} {Annalen der Physik}\ }\textbf
  {\bibinfo {volume} {17}},\ \bibinfo {pages} {132} (\bibinfo {year}
  {1900})}\BibitemShut {NoStop}%
\bibitem [{\citenamefont {Bohr}(1913)}]{Bohr}%
  \BibitemOpen
  \bibfield  {author} {\bibinfo {author} {\bibfnamefont {N.}~\bibnamefont
  {Bohr}},\ }\bibfield  {title} {\bibinfo {title} {On the constitution of atoms
  and molecules},\ }\href {https://doi.org/10.1080/14786441308634955}
  {\bibfield  {journal} {\bibinfo  {journal} {The London, Edinburgh, and Dublin
  Philosophical Magazine and Journal of Science}\ }\textbf {\bibinfo {volume}
  {26}},\ \bibinfo {pages} {1} (\bibinfo {year} {1913})},\ \Eprint
  {https://arxiv.org/abs/https://doi.org/10.1080/14786441308634955}
  {https://doi.org/10.1080/14786441308634955} \BibitemShut {NoStop}%
\bibitem [{\citenamefont {Heisenberg}(1925)}]{Heis1}%
  \BibitemOpen
  \bibfield  {author} {\bibinfo {author} {\bibfnamefont {W.}~\bibnamefont
  {Heisenberg}},\ }\bibfield  {title} {\bibinfo {title} {Über
  quantentheoretische umdeutung kinematischer und mechanischer beziehungen.},\
  }\href {https://doi.org/10.1007/BF01328377} {\bibfield  {journal} {\bibinfo
  {journal} {Zeitschrift für Physik}\ }\textbf {\bibinfo {volume} {33}},\
  \bibinfo {pages} {879–893} (\bibinfo {year} {1925})}\BibitemShut {NoStop}%
\bibitem [{\citenamefont {Born}\ and\ \citenamefont {Jordan}(1925)}]{BoJo}%
  \BibitemOpen
  \bibfield  {author} {\bibinfo {author} {\bibfnamefont {M.}~\bibnamefont
  {Born}}\ and\ \bibinfo {author} {\bibfnamefont {P.}~\bibnamefont {Jordan}},\
  }\bibfield  {title} {\bibinfo {title} {Zur quantenmechanik. ii.},\ }\href
  {https://doi.org/10.1007/BF01328531} {\bibfield  {journal} {\bibinfo
  {journal} {Zeitschrift für Physik}\ }\textbf {\bibinfo {volume} {34}},\
  \bibinfo {pages} {858–888} (\bibinfo {year} {1925})}\BibitemShut {NoStop}%
\bibitem [{\citenamefont {Born}\ \emph {et~al.}(1926)\citenamefont {Born},
  \citenamefont {Heisenberg},\ and\ \citenamefont {Jordan}}]{BoJoHe}%
  \BibitemOpen
  \bibfield  {author} {\bibinfo {author} {\bibfnamefont {M.}~\bibnamefont
  {Born}}, \bibinfo {author} {\bibfnamefont {W.}~\bibnamefont {Heisenberg}},\
  and\ \bibinfo {author} {\bibfnamefont {P.}~\bibnamefont {Jordan}},\
  }\bibfield  {title} {\bibinfo {title} {Zur quantenmechanik},\ }\href
  {https://doi.org/10.1007/BF01379806} {\bibfield  {journal} {\bibinfo
  {journal} {Zeitschrift für Physik}\ }\textbf {\bibinfo {volume} {35}},\
  \bibinfo {pages} {557–615} (\bibinfo {year} {1926})}\BibitemShut {NoStop}%
\bibitem [{\citenamefont {Gardner}(1970)}]{Gardner}%
  \BibitemOpen
  \bibfield  {author} {\bibinfo {author} {\bibfnamefont {M.}~\bibnamefont
  {Gardner}},\ }\bibfield  {title} {\bibinfo {title} {The fantastic
  combinations of john conway's new solitaire game 'life'},\ }\href@noop {}
  {\bibfield  {journal} {\bibinfo  {journal} {Scientific American}\ }\textbf
  {\bibinfo {volume} {223}},\ \bibinfo {pages} {120–123} (\bibinfo {year}
  {1970})}\BibitemShut {NoStop}%
\bibitem [{\citenamefont {Rovelli}(1996)}]{Rovelli}%
  \BibitemOpen
  \bibfield  {author} {\bibinfo {author} {\bibfnamefont {C.}~\bibnamefont
  {Rovelli}},\ }\bibfield  {title} {\bibinfo {title} {Relational quantum
  mechanics},\ }\href {https://doi.org/https://doi.org/10.1007/BF02302261}
  {\bibfield  {journal} {\bibinfo  {journal} {International Journal of
  Theoretical Physics}\ ,\ \bibinfo {pages} {1637}} (\bibinfo {year}
  {1996})}\BibitemShut {NoStop}%
\bibitem [{\citenamefont {Heisenberg}(1999)}]{Heis2}%
  \BibitemOpen
  \bibfield  {author} {\bibinfo {author} {\bibfnamefont {W.}~\bibnamefont
  {Heisenberg}},\ }\href@noop {} {\emph {\bibinfo {title} {Lo sfondo filosofico
  della fisica moderna}}}\ (\bibinfo  {publisher} {Sellerio},\ \bibinfo
  {address} {Palermo},\ \bibinfo {year} {1999})\BibitemShut {NoStop}%
\bibitem [{\citenamefont {Laudisa}(2019)}]{Laudisa}%
  \BibitemOpen
  \bibfield  {author} {\bibinfo {author} {\bibfnamefont {F.}~\bibnamefont
  {Laudisa}},\ }\href@noop {} {\emph {\bibinfo {title} {La realtà al tempo dei
  quanti. Einstein, Bohr e la nuova immagine del mondo}}}\ (\bibinfo
  {publisher} {Bollati Boringhieri},\ \bibinfo {address} {Torino},\ \bibinfo
  {year} {2019})\BibitemShut {NoStop}%
\bibitem [{\citenamefont {Abbott}\ \emph {et~al.}(2016)\citenamefont {Abbott},
  \citenamefont {Abbott}, \citenamefont {Abbott}, \citenamefont {Abernathy},\
  and\ \citenamefont {et~al.}}]{Ligo}%
  \BibitemOpen
  \bibfield  {author} {\bibinfo {author} {\bibfnamefont {B.~P.}\ \bibnamefont
  {Abbott}}, \bibinfo {author} {\bibfnamefont {R.}~\bibnamefont {Abbott}},
  \bibinfo {author} {\bibfnamefont {T.~D.}\ \bibnamefont {Abbott}}, \bibinfo
  {author} {\bibfnamefont {M.~R.}\ \bibnamefont {Abernathy}},\ and\ \bibinfo
  {author} {\bibnamefont {et~al.}} (\bibinfo {collaboration} {LIGO Scientific
  Collaboration and Virgo Collaboration}),\ }\bibfield  {title} {\bibinfo
  {title} {Observation of gravitational waves from a binary black hole
  merger},\ }\href {https://doi.org/10.1103/PhysRevLett.116.061102} {\bibfield
  {journal} {\bibinfo  {journal} {Phys. Rev. Lett.}\ }\textbf {\bibinfo
  {volume} {116}},\ \bibinfo {pages} {061102} (\bibinfo {year}
  {2016})}\BibitemShut {NoStop}%
\bibitem [{\citenamefont {Kant}(1976)}]{Kant}%
  \BibitemOpen
  \bibfield  {author} {\bibinfo {author} {\bibfnamefont {I.}~\bibnamefont
  {Kant}},\ }\href@noop {} {\emph {\bibinfo {title} {Critica della ragione
  pura}}}\ (\bibinfo  {publisher} {Adelphi},\ \bibinfo {address} {Milano},\
  \bibinfo {year} {1976})\BibitemShut {NoStop}%
\bibitem [{\citenamefont {Forrest}(2020)}]{Leibniz}%
  \BibitemOpen
  \bibfield  {author} {\bibinfo {author} {\bibfnamefont {P.}~\bibnamefont
  {Forrest}},\ }\bibfield  {title} {\bibinfo {title} {{The Identity of
  Indiscernibles}},\ }in\ \href@noop {} {\emph {\bibinfo {booktitle} {The
  {Stanford} Encyclopedia of Philosophy}}},\ \bibinfo {editor} {edited by\
  \bibinfo {editor} {\bibfnamefont {E.~N.}\ \bibnamefont {Zalta}}}\ (\bibinfo
  {publisher} {Metaphysics Research Lab, Stanford University},\ \bibinfo {year}
  {2020})\ \bibinfo {edition} {{W}inter 2020}\ ed.\BibitemShut {Stop}%
\bibitem [{\citenamefont {Cartesio}(1986)}]{Cartesio}%
  \BibitemOpen
  \bibfield  {author} {\bibinfo {author} {\bibnamefont {Cartesio}},\
  }\href@noop {} {\emph {\bibinfo {title} {Opere filosofiche}}}\ (\bibinfo
  {publisher} {Laterza},\ \bibinfo {address} {Roma},\ \bibinfo {year}
  {1986})\BibitemShut {NoStop}%
\bibitem [{\citenamefont {Penrose}(2018)}]{Penrose}%
  \BibitemOpen
  \bibfield  {author} {\bibinfo {author} {\bibfnamefont {R.}~\bibnamefont
  {Penrose}},\ }\href@noop {} {\emph {\bibinfo {title} {La strada che porta
  alla realtà. Le leggi fondamentali dell'universo}}}\ (\bibinfo  {publisher}
  {Rizzoli},\ \bibinfo {address} {Milano},\ \bibinfo {year} {2018})\BibitemShut
  {NoStop}%
\bibitem [{\citenamefont {Hawking}(2017)}]{Hawking}%
  \BibitemOpen
  \bibfield  {author} {\bibinfo {author} {\bibfnamefont {S.}~\bibnamefont
  {Hawking}},\ }\href@noop {} {\emph {\bibinfo {title} {The Grand Design}}}\
  (\bibinfo  {publisher} {Transworld},\ \bibinfo {year} {2017})\BibitemShut
  {NoStop}%
\bibitem [{\citenamefont {Landsman}(2021)}]{Landsman}%
  \BibitemOpen
  \bibfield  {author} {\bibinfo {author} {\bibfnamefont {K.}~\bibnamefont
  {Landsman}},\ }\bibfield  {title} {\bibinfo {title} {Indeterminism and
  undecidability},\ }\href {https://doi.org/10.1007/978-3-030-70354-7_3} {\ ,\
  \bibinfo {pages} {17} (\bibinfo {year} {2021})}\BibitemShut {NoStop}%
\bibitem [{\citenamefont {Poletti}(2023)}]{Poletti2}%
  \BibitemOpen
  \bibfield  {author} {\bibinfo {author} {\bibfnamefont {M.}~\bibnamefont
  {Poletti}},\ }\bibfield  {title} {\bibinfo {title} {On the strangeness of
  quantum probabilities},\ }\bibfield  {journal} {\bibinfo  {journal} {Quantum
  Studies: Mathematics and Foundations}\ }\href
  {https://doi.org/10.1007/s40509-023-00299-z} {10.1007/s40509-023-00299-z}
  (\bibinfo {year} {2023})\BibitemShut {NoStop}%
\bibitem [{\citenamefont {Schrodinger}(1935)}]{Shro}%
  \BibitemOpen
  \bibfield  {author} {\bibinfo {author} {\bibfnamefont {E.}~\bibnamefont
  {Schrodinger}},\ }\bibfield  {title} {\bibinfo {title} {{Die gegenwartige
  Situation in der Quantenmechanik}},\ }\href
  {https://doi.org/10.1007/BF01491891} {\bibfield  {journal} {\bibinfo
  {journal} {Naturwiss.}\ }\textbf {\bibinfo {volume} {23}},\ \bibinfo {pages}
  {807} (\bibinfo {year} {1935})}\BibitemShut {NoStop}%
\bibitem [{\citenamefont {Poletti}(2022)}]{Poletti1}%
  \BibitemOpen
  \bibfield  {author} {\bibinfo {author} {\bibfnamefont {M.}~\bibnamefont
  {Poletti}},\ }\bibfield  {title} {\bibinfo {title} {On the strangeness of
  quantum mechanics},\ }\href {https://doi.org/10.1007/s10701-022-00582-w}
  {\bibfield  {journal} {\bibinfo  {journal} {Foundations of Physics}\ }\textbf
  {\bibinfo {volume} {52}},\ \bibinfo {pages} {1} (\bibinfo {year}
  {2022})}\BibitemShut {NoStop}%
\bibitem [{\citenamefont {Everett~III}(1957)}]{Everett}%
  \BibitemOpen
  \bibfield  {author} {\bibinfo {author} {\bibfnamefont {H.}~\bibnamefont
  {Everett~III}},\ }\bibfield  {title} {\bibinfo {title} {"relative state"
  formulation of quantum mechanics},\ }\bibfield  {journal} {\bibinfo
  {journal} {Reviews of Modern Physics}\ }\textbf {\bibinfo {volume} {29.3}},\
  \href {https://doi.org/https://doi.org/10.1103/RevModPhys.29.454}
  {https://doi.org/10.1103/RevModPhys.29.454} (\bibinfo {year}
  {1957})\BibitemShut {NoStop}%
\bibitem [{\citenamefont {Vaidman}(2021)}]{Vaidman}%
  \BibitemOpen
  \bibfield  {author} {\bibinfo {author} {\bibfnamefont {L.}~\bibnamefont
  {Vaidman}},\ }\href@noop {} {\bibinfo {title} {Many-worlds interpretation of
  quantum mechanics}},\ \bibinfo {howpublished}
  {\url{https://plato.stanford.edu/archives/fall2021/entries/qm-manyworlds/}}
  (\bibinfo {year} {2021})\BibitemShut {NoStop}%
\bibitem [{\citenamefont {Wittgenstein}(1964)}]{Witg}%
  \BibitemOpen
  \bibfield  {author} {\bibinfo {author} {\bibfnamefont {L.}~\bibnamefont
  {Wittgenstein}},\ }\href@noop {} {\emph {\bibinfo {title} {Tractatus
  logico-philosophicus e Quaderni 1914-1916}}}\ (\bibinfo  {publisher}
  {Einaudi},\ \bibinfo {address} {Torino},\ \bibinfo {year} {1964})\BibitemShut
  {NoStop}%
\bibitem [{\citenamefont {Feynman}()}]{Fey}%
  \BibitemOpen
  \bibfield  {author} {\bibinfo {author} {\bibfnamefont {R.}~\bibnamefont
  {Feynman}},\ }\href@noop {} {\bibinfo {title} {Nobody understands quantum
  mechanics}}\BibitemShut {NoStop}%
\bibitem [{\citenamefont {Gödel}(1999)}]{Godel}%
  \BibitemOpen
  \bibfield  {author} {\bibinfo {author} {\bibfnamefont {K.}~\bibnamefont
  {Gödel}},\ }\href@noop {} {\emph {\bibinfo {title} {Opere, Volume 1
  1929-1936}}}\ (\bibinfo  {publisher} {Bollati Boringhieri},\ \bibinfo
  {address} {Torino},\ \bibinfo {year} {1999})\BibitemShut {NoStop}%
\bibitem [{\citenamefont {Wheeler}(1974)}]{Wheeler}%
  \BibitemOpen
  \bibfield  {author} {\bibinfo {author} {\bibfnamefont {J.}~\bibnamefont
  {Wheeler}},\ }\href@noop {} {\bibinfo {title} {Add “participant” to
  “undecidable propositions” to arrive at physics}},\ \bibinfo
  {howpublished} {https://jawarchive.files.wordpress.com/2012/03/twa-1974.pdf}
  (\bibinfo {year} {1974})\BibitemShut {NoStop}%
\bibitem [{\citenamefont {Szangolies}(2018)}]{Szangolies}%
  \BibitemOpen
  \bibfield  {author} {\bibinfo {author} {\bibfnamefont {J.}~\bibnamefont
  {Szangolies}},\ }\bibfield  {title} {\bibinfo {title} {Epistemic horizons and
  the foundations of quantum mechanics},\ }\href
  {https://doi.org/10.1007/s10701-018-0221-9} {\bibfield  {journal} {\bibinfo
  {journal} {Foundations of Physics}\ }\textbf {\bibinfo {volume} {48}},\
  \bibinfo {pages} {1669–} (\bibinfo {year} {2018})}\BibitemShut {NoStop}%
\bibitem [{\citenamefont {Brukner}(2009)}]{Brukner}%
  \BibitemOpen
  \bibfield  {author} {\bibinfo {author} {\bibfnamefont {{\v{C}}.}~\bibnamefont
  {Brukner}},\ }\bibfield  {title} {\bibinfo {title} {Quantum complementarity
  and logical indeterminacy},\ }\href@noop {} {\bibfield  {journal} {\bibinfo
  {journal} {Natural Computing}\ }\textbf {\bibinfo {volume} {8}},\ \bibinfo
  {pages} {449} (\bibinfo {year} {2009})}\BibitemShut {NoStop}%
\bibitem [{\citenamefont {Wigner}(1995)}]{Wigner}%
  \BibitemOpen
  \bibfield  {author} {\bibinfo {author} {\bibfnamefont {E.~P.}\ \bibnamefont
  {Wigner}},\ }\bibfield  {title} {\bibinfo {title} {Remarks on the mind-body
  question},\ }\href {https://doi.org/10.1007/978-3-642-78374-6_20} {\bibfield
  {journal} {\bibinfo  {journal} {Philosophical Reflections and Syntheses}\ ,\
  \bibinfo {pages} {247}} (\bibinfo {year} {1995})}\BibitemShut {NoStop}%
\end{thebibliography}%

\end{document}